\documentclass[11pt,dvips]{article}

\usepackage{epsfig,times} 
\usepackage{picinpar}
\usepackage{wrapfig}
\usepackage{floatflt}
\makeatletter
\renewcommand{\section}{\@startsection{section}{1}{0in}
	{0.4\baselineskip}{0.1\baselineskip}{\Large\bf}}
\renewcommand{\subsection}{\@startsection{subsection}{2}{0in}
	{0.25\baselineskip}{-\baselineskip}{\large\bf}}
\renewcommand{\subsubsection}{\@startsection{subsubsection}{3}{0in}
	{0.1\baselineskip}{-\baselineskip}{\normalsize\bf}}
\makeatother
\newcommand{\icrc}{$26^{\rm th}$ ICRC\ }
\begin{document}

%
\makeatletter\newcommand{\ps@icrc}{
\renewcommand{\@oddhead}{\slshape{OG.2.4.19}\hfil}}
\makeatother\thispagestyle{icrc}
%
%

\begin{center}
%
{\LARGE \bf The Physics Potential of Ground-Based Gamma-Ray Astronomy below 50 GeV}
\end{center}

\begin{center}
%
%
{\bf N. Magnussen$^{1}$}\\
{\it $^1$ Fachbereich Physik, Universit\"at Wuppertal, Gau\ss str. 20, 42097 Wuppertal, Germany}
\end{center}

\begin{center}
{\large \bf Abstract\\}
\end{center}
\vspace{-0.5ex}
%
%
Currently three ground-based Air Cherenkov detectors
with energy thresholds below 50 GeV are being commissioned
(CELESTE and STACEE) or under construction (MAGIC Telescope).
Based on the expected performance of the MAGIC Telescope
and with an emphasis on those physics questions
which are unique to the energy domain below 50 GeV
an overview of the scientific prospects of
ground-based high energy Gamma-Ray Astronomy in terms of astrophysics,
cosmology and particle physics questions is given.
%

\vspace{1ex}

%
%
\section{Introduction}
\label{intro.sec}

Technical developments have so far allowed  to
observe the universe from radio waves to $\gamma$-rays up to about 10~GeV and from
about 300~GeV up to 100~TeV.  A gap has remained between 10~GeV and 300~GeV which 
is going to be investigated with the MAGIC Telescope 
currently under construction (Martinez 1999).
This instrument will be an Imaging Air Cherenkov Telescope (IACT)
employing advanced technology for all of its ingredients. 
Other ground-based projects aiming at the energy domain below 100 GeV
are the solar array projects CELESTE and STACEE. 

\section{The Cosmological Gamma-Ray Horizon}
\label{cosmo.sec}

In spite of an energy-flux sensitivity superior to
the EGRET instrument onboard
CGRO (for energy spectra extrapolated to higher energies),
a much smaller number of sources (e.g. 2 confirmed and 3 unconfirmed extragalactic
sources) has been discovered with the IACT technique above 300~GeV implying that most
of the EGRET sources have spectra
turning over between 10~GeV and 300~GeV. For extragalactic sources
this might either be due to external absorption
on the diffuse cosmological background or due to internal absorption in the sources.

That above some critical energy defining the $\gamma$-ray horizon
the visible universe in high energy photons should be limited because of pair
production on the cosmological low-energy diffuse background photons was first pointed
out by Gould \& Schr\'eder (1966).
With increasing $\gamma$-ray energy, the threshold condition is fulfilled
for an increasing number of low-energy photons from
the diffuse radiation background, resulting in a
decreasing $\gamma$-ray horizon.  
Conversely, triggering at $\gamma$-ray energies
lower than current IACTs can observe one will have access to a much larger fraction of
the Hubble volume and thus to a much larger source population.  
According to the current best estimates of the diffuse background current IACTs
should only be able to observe the universe out to redshifts of z $ \approx 0.1$.

With the MAGIC Telescope operating above 10 GeV a large number of AGN
will be observable, and a population study of all results
should then reveal a plot of the highest energy seen by the MAGIC Telescope, vs
redshift. The slope of the locus of energy maxima should then be
proportional to the density of intergalactic target photons. However,
those maximum energies which are below the locus of points, should be due
to intrinsic source absorption, and can be used to constrain $\gamma$-ray
production models. 

\begin{figure}[t]
\vspace{-5cm}
\begin{center}
\epsfig{figure=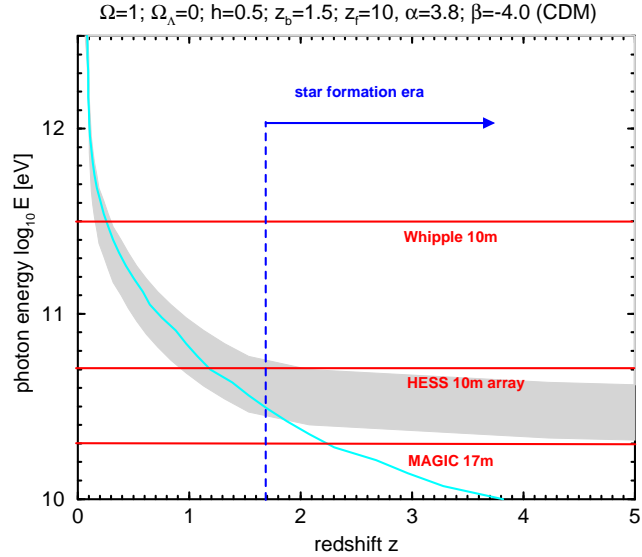,height=6.1in}
\end{center}
\vspace{-1cm}
\caption{The shape of the $\gamma$-ray horizon resulting from a calculation by
Mannheim (1999). The grey band indicates the uncertainty
in the calculation due to uncertainty in the diffuse photon background level.
The line is the prediction based on an extreme assumption
of the star formation rate continuing at the maximum level beyond $z$ = 1.5
and negligible dust absorption.
For distant sources only $\gamma$-rays with energies less than the 
horizon can reach the observer. 
Taken from Mannheim (1999).}
\label{fig:horizon}
\end{figure}

The flux of the isotropic radiation background from the far-infrared to 
the ultra-violet is only poorly known from direct measurements.
The measurement of turn-over energies in the spectra of 
extragalactic sources will allow to infer
the low-energy background flux in a manner completely independent of conventional methods.
The background radiation flux is an important observable for models of cosmic
structure formation because it constitutes a convolution
of the star formation history, the dust extinction history, and the
evolution of the initial mass function. In addition
it has some sensitivity to the existence of massive neutrinos or stable
particles from supersymmetric extensions of the Standard Model acting as dark
matter. In Fig.~\ref{fig:horizon} the shape of the cosmological 
$\gamma$-ray horizon 
as calculated by Mannheim (1999) is shown. Also shown 
as horizontal lines are the lowest energies
measurable by the Whipple telescope and the planned HESS array of 16 IACTs and the
MAGIC Telescope in phase 1 (i.e. equipped with classical PMTs). 
As AGN activity seems to be linked to galaxy merger activity and the star formation
era the volume density of AGN shows a prominent peak or
plateau at $z$ $\geq$ 1. 
Only the MAGIC Telescope will thus have access to the bulk of cosmological AGN
and it will be the only IACT with access to $\gamma$-rays
with energies below the (possibly) asymptotic $\gamma$-ray horizon.
The exact shape of the $\gamma$-ray horizon will also
depend on the cosmological
parameters and constitutes an important cross-check of their values determined
by other means. 

\section{Resolving the Diffuse Gamma-Ray Background}
\label{diffuse.sec}

An important scientific question closely connected to the asymptotic $\gamma$-ray horizon
is the understanding of the diffuse $\gamma$-ray background in the GeV energy
domain in terms of contributions from point and diffuse sources. The only known
point sources today are AGN. An analysis of the AGN $\gamma$-ray
luminosity function based on the EGRET results by Chiang \& Mukherjee (1998)
indicates that a significant fraction of the diffuse $\gamma$-ray background 
may be due to diffuse sources. 

The emissivity of the universe at energies above
the energy of the asymptotic horizon on Hubble length and Hubble time
scales will appear in the diffuse $\gamma$-ray background at energies {\it below}
the asymptotic value due to the pathlength for the cascading
process (initiated at high energies due to interactions 
with the diffuse background)
reaching a length scale of the order of the Hubble length.
This also stresses
the importance of the location and shape of the asymptotic
$\gamma$-ray horizon.
The determination of the AGN luminosity function from
high sensitivity data as will be provided by the MAGIC Telescope
thus is of fundamental importance for the understanding of the high
energy emissivity.

Possible diffuse sources which could inject significant amounts of energy
into the universe at high energies are 
the decays of supermassive relics from the earliest epochs, e.g.
Higgs bosons in certain realizations of supersymmetric extensions of the
Standard Model and topological defects.

\section{Gamma-Ray Emission from Pulsars}
\label{pulsar1.sec}

Of the more than 800 known radio pulsars EGRET has revealed 7 to emit pulsed
$\gamma$-rays up to $\approx$ 10~GeV. 
No steady pulsed emission from pulsars has yet been
detected by ground-based IACTs above 300 GeV.
To clarify the production mechanism,
measurements in the 10~GeV to 100~GeV energy domain are crucial.
The polar cap model for pulsed emission (Harding 1981)
predicts a sharp
cutoff in the $\gamma$-ray spectra above a few GeV due to
absorption in the strong magnetic field. As
detailed phase resolved modelling
showed, the bridging emission between the two
pulses should exhibit harder spectra (Daugherty \& Harding 1996).
This prediction was recently confirmed by
phase resolved spectroscopy of the Crab, Vela and Geminga pulsars
(Fierro et al. 1998) and provides a
very low threshold IACT like the MAGIC Telescope with the unique
opportunity to provide answers on the emission regions for the
highest energy $\gamma$-rays from the neutron star magnetosphere.
Phase resolved rates above 10 GeV of up to more than
100$ \sigma$ per hour per 0.1 phase interval (after
image analysis), or more than $\sim 10 \sigma$ if no background cuts are made,
are estimated based on the MAGIC Telescope's sensitivity.
Note that the locking onto the pulsed signal will be easily achieved
and thus suppress any systematic background effects.


\section{Pulsar Discovery Potential of the MAGIC Telescope}
\label{pulsar2.sec}

Besides the known $\gamma$-ray pulsars there is a long list
of unidentified EGRET sources which seem to be coincident with
massive star forming OB stellar associations (Yadigaroglu \& Romani
1997).
Whereas the radio signals are absorbed by the relatively high
electron densities inside the OB regions, the $\gamma$-rays should escape
unabsorbed.
The large photon collection area of the MAGIC Telescope
will allow to detect new pulsars with pulsed fluxes at 20 GeV
in the order of 10$^{-14}$ to 10$^{-13}$ cm$^{-2}$s$^{-1}$MeV$^{-1}$
on time scales between a few hours and a few minutes.
EGRET-quiet radio pulsars should also be monitored as the flux above
100 MeV may be too small to be detectable by EGRET, but stronger than
10$^{-14}$ cm$^{-2}$s$^{-1}$MeV$^{-1}$ at 20 GeV. Detections of
this type of pulsar would
provide valuable data for the dependence of $\gamma$-ray
luminosities on basic pulsar quantities (de Jager 1998).

\section{Gamma-Ray Supernova Remnants}
\label{snr.sec}

Supernova remnants (SNRs), possible
sites of cosmic ray acceleration favoured in most
models of the cosmic ray origin, seem to be more complex than previously
believed (Jones 1997). Although four SNRs have been observed above 300~GeV
(Crab nebula, Vela, PSR1706-44, and SN1006), the question of the origin of cosmic
rays
is far from answered. More sensitive measurements at lower energies will be
of great importance in identifying the spectral component showing up
above 300~GeV in the four sources above and to discover $\gamma$-ray
emission in more SNRs. With the
low energy threshold
of the MAGIC Telescope it may be possible to observe a {\it two component $\gamma$-ray
spectrum}, which should then allow to decouple the
predicted leptonic and
hadronic components in SNR shells.

\section{Search for a Cold Dark Matter Candidate}
\label{cdm.sec}

Among the candidates for the dark matter in the universe, weakly interacting
massive particles (WIMPs) such as the neutralinos arising in supersymmetric
extensions of the standard model are the most plausible.  Their interaction
cross section and expected mass naturally match to produce the dominant
contribution to the energy density of an expanding Friedmann-Robertson universe.
With a lower limit on their mass from particle physics of about 20 GeV,
the interesting mass range of $\approx$20~GeV to 300~GeV implies that $\gamma$-rays
from decay or annihilation (neutralinos are Majorana particles) could be
discovered with the MAGIC Telescope, e.g.  as a $\gamma$-ray line from the
region of the Galactic Centre. Note that from the northern site (La
Palma) the effective photon collection area will be of the order of
10$^6$ m$^2$ which will yield sufficient sensitivity to cover
a fair fraction of the
MSSM parameter space (see e.g. Bergstrom et al. 1998).

\section{High-Energy Counterparts of Gamma-Ray Bursts}
\label{grb.sec}

The low moment of inertia is one of the main features of the MAGIC Telescope
and it will allow rapid positioning
towards observation targets (typically within
30 s).  The telescope is thus ideally suited to search for high-energy
counterparts of GRBs.  The low-energy threshold will allow
observations out to large cosmological distances.  Extrapolation of the
energy spectra of the GRBs
detected by EGRET
leads to the prediction
that even medium-strength bursts will yield very high
$\gamma$ rates detectable by the MAGIC Telescope ($\sim $~kHz)
due to the very large effective collection area.  For $\gamma$
rates of this magnitude the 
MAGIC Telescope (in phase 2) will measure energy spectra from
about 5 GeV up to the highest energies.

\section{Probing the Quantum Gravity Scale}
\label{qg.sec}

Because of the high rates expected for GRBs observed
with the MAGIC Telescope the high energy light\-curve
of GRBs can be obtained with good temporal resolution. A number of Quantum Gravity
(QG) models (Amelino-Camelia et al. 1997) predict modified laws of propagation of neutral
particles as a result of interactions with the quantum gravity medium.
The time delays due to the ensuing dispersion relation are only then
significantly larger than the Planck time when 
the particles of differing energies have either energies close
to the QG scale (which is expected to be close to the Planck scale, i.e.,10$^{19}$ GeV)
or have traversed cosmological distances. This last requirement
is fulfilled for at least a subclass of GRB which have been observed
to have redshifts up to more than 3.4. With an
assumed time resolution for the lightcurve
of 1 sec and a pathlength of more than several Gpc, the sensitivity
of MAGIC Telescope's measurements for the QG scale
will be of the order of the Planck scale. For comparison, 
the current best limits are about 1\% of the Planck scale.
Recent work within the Liouville string formulation of QG
(Ellis et al. 1999) yields a refractive index which increases linearly with
the photon energy, i.e. the high energy photons will arrive {\it later}
compared to the low energy photons. This distinctive signature will 
help distinguishing the QG effect from classical dispersion effects which
yield increasing time delays for decreasing energies.

\vspace{1ex}
\begin{center}
{\Large\bf References}
\end{center}

\noindent
Amelino-Camelia G., Ellis J., Mavromatos N.\,E. \& Nanopoulos D.\,V.
1997, Mod.\,Phys. A12, 607 \\
Bergstrom L., Ullio P. \& Buckley J.\,H. 1998, Astropart. Phys. 9, 137
\\
Chiang J. \& Mukherjee R. 1998, ApJ 496, 752 \\
Daugherty J.\,K. \& Harding A.\,K. 1996, ApJ 458, 278 \\
Ellis J., Mavromates N.E. \& Nanoloulos D.V. 1999, gr-qc/9904068 \\
de Jager O.\,C. 1998, Proc.\, 26th European Cosmic Ray Conference,
Madrid, Spain, July 1998 \\
Fierro J.\,M. et al. 1998, ApJ 494, 734 \\
Gould R.\,J. \& Schr\'eder G., 1966, Phys.\,Rev.\,Lett. 16, 252 \\
Harding A.\,K. 1981 , ApJ 245, 267 \\
Jones T.\,W. 1997, astro-ph/9710227 \\
Mannheim K. 1999, Rev.\,Mod.\,Astrophy. 12, 101 \\
Martinez M. 1999, Proc. 26th ICRC (Salt Lake City), OG.4.3.08 \\
Yadigaroglu I.-A. \& Romani R.\,W. 1997, ApJ 476, 347
\end{document}